\documentclass[aps,prl,reprint,ttisuperscriptaddress,twocolumn]{revtex4}
\usepackage[dvips]{graphicx}
\usepackage{subfigure}
\usepackage{bm}
\usepackage{color}
\usepackage{amsmath,amssymb}
\begin{document}
\title{
Electric voltage generation by antiferromagnetic dynamics 
}
\author{Yuta Yamane$^1$, Jun'ichi Ieda$^{1,2}$, and Jairo Sinova$^{1,3}$}
\affiliation{$^1$Institut f\"{u}r Physik, Johannes Gutenberg Universit\"{a}t Mainz,D-55099 Mainz, Germany}
\affiliation{$^2$Advanced Science Research Center, Japan Atomic Energy Agency, Tokai, Ibaraki 319-1195, Japan}
\affiliation{$^3$Institute of Physics ASCR, v.v.i., Cukrovarnicka 10, 162 53 Praha 6, Czech Republic}
\date{\today}
\begin{abstract}
We theoretically demonstrate dc and ac electric voltage generation due to spinmotive forces originating from domain wall motion and magnetic resonance, respectively, in two-sublattice antiferromagnets.
Our theory accounts for the canting between the sublattice magnetizations, the nonadiabatic electron spin dynamics, and the Rashba spin-orbit coupling, with the inter-sublattice electron dynamics treated as a perturbation.
This work suggests a new way to observe and explore the dynamics of antiferromagnetic textures by electrical means, an important aspect in the emerging field of antiferromagnetic spintronics, where both manipulation and detection of antiferromagnets are needed.
\end{abstract}
\maketitle
%=========================================================================
%           Introduction
%=========================================================================
\emph{Introduction.---}
In magnetic materials, the exchange interaction between the conduction electron spin and the local magnetization is responsible for a variety of important phenomena.
Among the spintronic effects caused by this interaction, spin-transfer torque provides a path to promising information technology by enabling the angular momentum to be transferred between the electrons and magnetization\cite{sloncz,berger}.
The same interaction can also mediate a transfer of energies between the two channels.
Such a transfer is mediated by the spinmotive force (SMF)\cite{berger-smf,stern,barnes,review}, where magnetic energies stored by the magnetization can be transformed to an electric voltage.
Theoretically, the SMF is attributed to a spin-dependent electric field arising due to the exchange interaction\cite{korenman,volovik,duine,tserkov,yamane-jap,kim,tatara,yamane-prb}, termed spin electric field.
The SMF reflects the temporal and spatial variations of the magnetization, and thus offers a powerful way of probing and exploring the dynamics and nature of various magnetic textures.

The concept of SMF has been discussed for nonmagnetic materials under nonuniform magnetic fields\cite{stern}, for ferromagnets (FMs)\cite{berger-smf,barnes,review,korenman,volovik,duine,tserkov,yamane-jap}, and recently extended to FMs that are subject to the Rashba spin-orbit coupling (RSOC)\cite{kim,tatara,yamane-prb}.
The experiments that have observed SMF so far are all done in FMs\cite{yang,hai,yamane-prl,hayashi,tanabe}.
A question naturally arises;
can one expect SMF in materials with different magnetic orderings from FM, such as antiferromagnets (AFMs)\cite{cheng,okabayashi,gomonay}?
AFMs are generating more attention in the field of spintronics due to their potential to become a key player in technological applications where AFMs play an active role\cite{tomas}.
This motivates the demand for reliable methods to observe dynamical AFM textures that are often difficult to see directly by conventional FM methods because of their small magnetization.
SMF, if present, would allow for a detection of the AFM dynamics by electrical means.

Recently, Cheng and Niu\cite{cheng} formulated a theory of electron dynamics in two-sublattice AFMs, discussing Berry's phase effects.
One of their predictions is that no electric voltage appears unless a nonequilibrium spin polarization is generated by externally injecting spin into the AFM.
Such a prediction was underpinned, however, by the assumption of perfect collinearity between the sublattice magnetizations, the half-metallic nature in each sublattice, and the absence of RSOC.

In this work, we predict a finite electric voltage to appear in two-sublattice AMFs without the need to apply an external spin source, when one relaxes the conditions imposed in the previous work;
i.e., one allows for i) the canting of the sublattice magnetizations, ii) the electron spin-flip process within each sublattice, and iii) the presence of RSOC.
We formulate the SMF, deriving expressions of the spin electric and magnetic fields in the case where the exchange coupling is so large that the inter-sublattice electron dynamics can be treated perturbativtely.
We then demonstrate that domain wall (DW) motion and antiferromagnetic resonance (AFMR) generate dc and ac SMFs, respectively.
These results indicate the capacity of SMF for detecting AFM dynamics by generating electric signals.

%=========================================================================
%           Model
%=========================================================================
\emph{Model.---}
We consider an AFM metal composed of two sublattices (1 and 2) with equal saturation magnetization $M_{\rm S}$.
In order to treat the magnetization classically, the coarse graining for the magnetic channel is performed.
The classical and continuous vector ${\bm m}_1 ({\bm r},t)$ $(|{\bm m}_1({\bm r},t)|=1)$ represents the direction of local magnetization in the sublattice 1, and the similar definition for ${\bm m}_2({\bm r},t)$;
here the lattice structure is smeared out and the magnetizations of both sublattices are defined at every point in space.
This classical treatment is justified when the spatial variation of the magnetization is sufficiently slow compared to the atomistic length scale.

For the conduction electron channel we assume the following four-band Hamiltonian;
\begin{eqnarray}
{\cal H} &=&   \left( \begin{array}{cc}  J{\bm \sigma}\cdot{\bm m}_1 ({\bm r},t) &  0 \\
 0 &  J{\bm \sigma}\cdot{\bm m}_2 ({\bm r},t)  \end{array} \right) \nonumber\\
 &&
+ \left( \begin{array}{cc}  t_{11}({\bm p}) &  t_{12}({\bm p}) \\
 t_{21}({\bm p}) &  t_{22} ({\bm p}) \end{array} \right)
 + \left( \begin{array}{cc}  {\cal H}_{\rm R1} &  0 \\
 0 &  {\cal H}_{\rm R2} \end{array} \right) ,
\label{h} \end{eqnarray}
where the upper-left (bottom-right) bands correspond to the sublattice 1 (2).
The first term describes the exchange interaction with $J$ being the exchange coupling energy and ${\bm \sigma}$ the Pauli matrices indicating the electron spin operator.
The second term is the kinetic energy tensor with ${\bm p}$ the momentum operator of the electron;
the diagonal and off-diagonal components describe the intra- and inter-sublttice electron dynamics, respectively, where we assume $t_{11}=t_{22}$ and $t_{12}=t^\dagger_{21}$.
The third term represents the RSOC, where ${\cal H}_{{\rm R}i}=(\lambda_i/\hbar){\bm \sigma}\cdot{\bm p}\times\hat{{\bm z}}$ $(i=1,2)$.
It was recently predicted that the Rashba constant $\lambda_i$ can be sublattice-dependent\cite{jakob,wadley}.

In this work, we assume that $t_{12}$ is a constant c-number and focus on the parameter regime of $|t_{12}|/J\ll1$\cite{note4}.
The inter-sublattice band-mixing, which is accompanied by the energy gain $\sim |t_{12}|$ as well as the energy cost $\sim2J$, can be thus treated as a perturbation.
This may be the case in, e.g., layered AMFs such as Mn$_2$Au\cite{jakob} and CuMnAs\cite{wadley} where the nearest-neighbor sites to hop are {\it intra}-sublattice with the c axis being the longest.

To expand ${\cal H}$ in powers of $J^{-1}$, let us perform the unitary transformation
\begin{equation}
{\cal H}' \equiv e^S \left( {\cal H} + i\hbar\partial_t  \right) e^{-S} ,
\label{h'}\end{equation}
with
\begin{equation}
S=  \frac{ {\bm \sigma} \cdot  {\bm n}}{2J} \left( \begin{array}{cc} 0 & t_{12}  \\  - t^*_{12} & 0 \end{array} \right) ,
\end{equation}
where ${\bm n}=({\bm m}_1-{\bm m}_2)/2$.
The AFM exchange coupling between ${\bm m}_1$ and ${\bm m}_2$ is mostly so large that $|{\bm n}|\simeq1$ and $|{\bm m}|\ll1$, where ${\bm m}=({\bm m}_1+{\bm m}_2)/2$.
The explicit expression of Eq.~(\ref{h'}) is written by
\begin{equation}
{\cal H}' = \left( \begin{array}{cc} {\cal H}_1 & t_{12} L \\  -t^*_{12} L & {\cal H}_2 \end{array} \right)  +
\left[ S, \left( \begin{array}{cc}  {\cal H}_{\rm R1} &  0 \\
 0 &  {\cal H}_{\rm R2} \end{array} \right) \right]
 + {\cal O}(J^{-2}) ,
\label{h'2}\end{equation}
with
\begin{eqnarray}
{\cal H}_i &=& \frac{{\bm p}^2}{2m_{\rm e}}  + (-1)^{i+1} J \left( 1 + \frac{ |t_{12}|^2}{2J^2} \right) {\bm \sigma}\cdot {\bm n} \nonumber\\
&& + J \left( 1 - \frac{ |t_{12}|^2}{2J^2} \right) {\bm \sigma}\cdot {\bm m}  + {\cal H}_{{\rm R}i} \nonumber\\
&\simeq& \frac{{\bm p}^2}{2m_{\rm e}} + J' {\bm \sigma} \cdot {\bm m}_i + {\cal H}_{{\rm R}i} , \quad (i=1,2) ,
\label{h_d}\end{eqnarray}
and
\begin{equation}
L = i \left[ \frac{\hbar}{2J} \left\{ ({\bm \sigma}\cdot\nabla{\bm n}) \cdot \frac{{\bm p}}{m_{\rm e}} + {\bm \sigma} \cdot \partial_t {\bm n} \right\} + {\bm \sigma}\cdot ({\bm n}\times{\bm m}) \right]  ,
\label{off} \end{equation}
where $J' = J(1+ |t_{12}|^2/2J^2)$, and we have assumed the quadratic dispersion $t_{11}= {\bm p}^2/2m_{\rm e}$ for the intra-sublattice kinetic energy.
Eq.~(\ref{h'2}) shows that, in the rotated frame, up to the order of $J^{-1}$ the off-diagonal components require the spatial and temporal variations in ${\bm n}$, the sublattice canting ${\bm m}$, and the RSOC\cite{note}.
$|{\bm m}|\ll1$ is used in the second equality in Eq.~(\ref{h_d}).

Now we assume the smooth and slow variations of the magnetizations by $\hbar|\partial_t{\bm n}|\ll  |t_{12}|$ and $\hbar |({\bm v}_F \cdot \nabla) {\bm n}|\ll |t_{12}| $ with ${\bm v}_F$ being the Fermi velocity of the conduction electrons, the small canting $|{\bm m}|\ll  | t_{12} | /J$, and $\lambda_i |{\bm p}_{\rm F}|/\hbar \ll  |t_{12}| $, so that the off-diagonal terms in Eq.~(\ref{h'2}) can be neglected compared to the diagonal terms.
Setting $10 |t_{12}|=J\sim1$ eV, the above conditions are usually well satisfied.

At this stage, in terms of the electron-magnetization interaction, it has been proven that the AFM can be treated as if it is two decoupled FMs;
the electrons couple to the magnetizations ${\bm m}_1$ and ${\bm m}_2$ in each sublattice (in the rotated frame) with the renormalized exchange coupling $J'$.
Eq.~(\ref{h_d}) is indeed the same form as the Hamiltonian commonly used for FMs.
We can thus derive spin electric and magnetic fields in each sublattice borrowing the theories for FMs\cite{berger-smf,volovik,barnes,review,korenman,duine,tserkov,yamane-jap,kim,tatara,yamane-prb}.
In doing so, an adiabatic approximation for the electron spin dynamics is adopted, i.e., assume $\tau_{\rm ex}^{-1}\gg\tau_{\rm sf}^{-1}$,  with $\tau_{\rm ex}=\hbar/2J'$ and $\tau_{\rm sf}$ being the relaxation time for the electron spin flip.
This condition ensures that the two band model with majority (almost antiparallel to ${\bm m}_i$) and minority (almost parallel to ${\bm m}_i$) spins is a good model in each sublattice.

The Hamiltonian (\ref{h'2}) with the above assumptions leads to the spin electric and magnetic fields\cite{berger-smf,volovik,barnes,review,korenman,duine,tserkov,yamane-jap,kim,tatara,yamane-prb},
\begin{eqnarray}
\bm{{\cal E}}_\pm^i &=& \pm \frac{\hbar}{2e} \left( {\bm m}_i \times \frac{\partial {\bm m}_i}{\partial t}  + \beta \frac{\partial {\bm m}_i}{\partial t}  \right) \cdot \nabla {\bm m}_i \nonumber \\
&& \pm \frac{\lambda_i m_{\rm e}}{e\hbar} \hat{{\bm z}} \times \frac{\partial{\bm m}_i}{\partial t}  ,
\label{e}\end{eqnarray}
\begin{eqnarray}
{\cal B}_{\pm, l}^i &=& \mp \epsilon_{lmn} \frac{\hbar}{4e} {\bm m}_i \cdot \frac{\partial{\bm m}_i}{\partial x_m} \times \frac{\partial{\bm m}_i}{\partial x_n} \nonumber\\
&&+ \frac{\lambda_i m_{\rm e}}{e\hbar} \left [\nabla \times \left( {\bm m} \times \hat{{\bm z}} \right) \right]_l
 \ \  (l,m,n=x,y,z),
\label{b}\end{eqnarray}
up to the first order of $\lambda_i$ and $\beta=\tau_{\rm ex}/\tau_{\rm sf}$.
Here $\epsilon_{lmn}$ is the Levi-Civita symbol.
Notice that the upper (lower) signs correspond to the majority (minority) electrons in each sublattice;
the spin fields operate on the majority and minority electrons in the opposite way, driving spin and charge currents within each sublattice.
The spin fields (\ref{e}) and (\ref{b}) can be responsible for, respectively, SMF and topological Hall effect.
Hereafter, we focus on effects of the spin electric field (\ref{e}).

$\bm{{\cal E}}_\pm^i$ gives rise to the charge current density ${\bm j}^i_{\rm c}=\left( \sigma_+ - \sigma_- \right) \bm{{\cal E}}^i_+$, where the electric conductivity $\sigma_{+(-)}$ for the majority (minority) electrons is assumed to be sublattice-independent, and $\bm{{\cal E}}_-^i=-\bm{{\cal E}}_+^i$ is used.
The sublatice-independency of the conductivities reflects the assumption that half the number of total electrons are accommodated in each sublattice due to the absence of external spin source.
${\bm j}^1_{\rm c}$ and ${\bm j}^2_{\rm c}$ are combined to induce the net charge current density ${\bm j}_{\rm c}= \sum_i {\bm j}^i_{\rm c}$.
When the system is an open circuit, the (ordinary) electric field ${\bm E}=-\nabla\phi-\partial{\bm A}/\partial t$ must also be present to cancel ${\bm j}_{\rm c}$, i.e., ${\bm j}_{\rm c} + 2(\sigma_++\sigma_-){\bm E}=0$, where $\phi$ and ${\bm A}$ are the electromagnetic potentials.
Adopting the Coulomb gauge, $\nabla\cdot{\bm A}=0$, the electric voltage $V$ between two given points ${\bm r}_a$ and ${\bm r}_b$ is obtained by
\begin{equation}
V = \int_{{\bm r}_a}^{{\bm r}_b} d{\bm r} \cdot \nabla \phi =  \frac{P}{2} \int_{{\bm r}_a}^{{\bm r}_b} d{\bm r} \cdot \sum_i \bm{{\cal E}}_+^i ,
\label{v}\end{equation}
with the spin polarization $P$ defined by $P=(\sigma_+ - \sigma_-)/(\sigma_+ + \sigma_-)$.

$\sum_i\bm{{\cal E}}_+^i$ is expressed in terms of ${\bm n}$ and ${\bm m}$ as
\begin{eqnarray}
\frac{1}{2}\sum_i \bm{{\cal E}}_+^i &=& \frac{\hbar\beta}{2e} \frac{\partial {\bm n}}{\partial t} \cdot \nabla {\bm n} + \frac{\left( \lambda_1 - \lambda_2 \right) m_{\rm e}}{2e\hbar} \hat{{\bm z}} \times \frac{\partial{\bm n}}{\partial t}  \nonumber \\
&& + \frac{\left( \lambda_1 + \lambda_2 \right) m_{\rm e}}{2e\hbar} \hat{{\bm z}} \times \frac{\partial{\bm m}}{\partial t}
+ {\bm f} ({\bm n},{\bm m}) .
\label{e_sum}\end{eqnarray}
In Eq.~(\ref{e_sum}) we have explicitly written out only three terms that will be relevant later when we demonstrate SMFs induced by DW motion and AFMR;
all the zeroth order terms in ${\bm m}$, which are the first two terms, and the third term that is one of the first order terms in ${\bm m}$.
The other terms, e.g., $(\hbar\beta/2e)\partial{\bm m}/\partial t \cdot\nabla{\bm m}$ and $(\hbar/2e){\bm m} \times \partial{\bm m}/\partial t \cdot\nabla{\bm m}$, the latter of which originates purely from the magnetization canting, are by definition all contained in ${\bm f}({\bm m},{\bm n})$.
Notice that the first term in Eq.~(\ref{e}) does not contribute to the net electron dynamics at the zeroth order in ${\bm m}$;
therefore, when ${\bm m}=0$, $\beta=0$ and $\lambda_i=0$, Eq.~(\ref{e_sum}) vanishes and there appears no electromotive force, which is consistent with the results in Ref.~\cite{cheng}.
Within the present framework, $\beta$ is the only quantity that explicitly reflects the renormalization $J\rightarrow J'$.
In general, ${\bm m}$ acquires finite magnitude due to external magnetic fields, Dzyaloshinskii-Moriya (DM) interaction, and temporal and spatial variations in the magnetizations\cite{ivanov,andreev,papa}.

Eq.~(\ref{e_sum}) is our key result.
In the following, we apply Eqs.~(\ref{v}) and (\ref{e_sum}) to the two systems:
DW motion and AFMR.

%=========================================================================
%          DW Motion
%=========================================================================
\emph{Domain wall motion.---}
The dynamics of magnetic DWs has been a central subject of study in the spintronics.
The early experimental confirmations of SMF were possible indeed by employing DW motion in FM permalloy nanowires\cite{yang,hayashi}.
Here we investigate the SMF induced by DW motion in AFM, assuming no RSOC for the moment;
the relevant term in Eq.~(\ref{e_sum}) up to the zeroth-order of ${\bm m}$ is the first term.

Consider a one-dimensional AFM nanowire with magnetic energy density $u=-2\mu_0 {\bm m} \cdot {\bm H} + A_0 {\bm m}^2 + A_1 \left( \partial{\bm n}/\partial z \right)^2 - K ( n_z^2+m_z^2 ) + D \hat{{\bm y}}\cdot ( {\bm n}\times {\bm m} ) $\cite{ivanov,andreev,bary} (see Fig.~1~(a) for the coordinate system).
The parameters $A_0$, $A_1$, $K$ and $D$ characterize the ``on-site'' and neighboring exchange couplings, the uniaxial anisotropy, and the DM interaction, respectively.
An equilibrium AFM texture is given by ${\bm n}$ and ${\bm m}$ at which $u$ takes an extremal value.
In the absence of external field, a static DW solution satisfying the boundary condition $n_z(\pm\infty)=\mp1$ is $\theta=2\tan^{-1} [e^{(z-q)/\Delta} ]$ and $\varphi=0$;
the polar angles are defined by ${\bm n}=(\sin\theta\cos\varphi,\sin\theta\sin\varphi,\cos\theta)$, $q$ represents the DW center position, and $\Delta=\sqrt{A_1/K}$.
The net magnetization ${\bm m}$ also forms the DW by ${\bm m}=(D/4A_0){\bm n}\times \hat{{\bm y}}$ due to the DM interaction.
Fig.~1~(a) shows a schematic of the one-dimensional DW.

This DW is driven into motion by magnetic field $H$ applied in the $x$ axis, developing the domain where ${\bm m}$ is parallel to the field [Fig.~1~(a)].
We assume that the dynamics of magnetizations ${\bm m}_1$ and ${\bm m}_2$ obeys the coupled Landau-Lifshitz-Gilbert equations,
\begin{equation}
\frac{\partial{\bm m}_i}{\partial t} =  \gamma {\bm m}_i \times \frac{\delta u}{\delta {\bm m}_i} + \alpha {\bm m}_i \times \frac{\partial{\bm m}_i}{\partial t} , \quad (i=1,2) ,
\label{llg}\end{equation}
where $\gamma$ is the gyromagnetic ratio and $\alpha$ is the Gilbert damping constant, both of which are assumed for simplicity to be sublattice-independent.
To obtain an analytical solution for the DW dynamics, we make the steady-motion approximation, where the DW maintains the equilibrium profile with $q$ and $\varphi$ being time dependent;
the DW dynamics is described by time evolution of the collective coordinates $(q,\varphi)$.
As it is known, the dynamics of AFM textures in general has an inertia\cite{ivanov,andreev}.
Applying the steady-motion approximation to Eq.~(\ref{llg}), the terminal velocity $v_{\rm DW}\equiv dq/dt |_{t=\infty}$ of the DW is obtained as\cite{bary}
\begin{equation}
v_{\rm DW} \simeq - \frac{D \Delta}{4\alpha\mu_0 M_{\rm S}H_E} \gamma H ,
\label{velocity}\end{equation}
where $H_E=A_0/\mu_0 M_{\rm S}$.
The azimuthal angle saturates at $\varphi (t=\infty)\simeq (\pi H/2\gamma\Delta) ( H^2 - D^2/4\mu_0^2 M_{\rm S}^2)^{-1} v_{\rm DW}$, the explicit value of which turns out to be irrelevant to the SMF.

\begin{figure}
\centering
\includegraphics[width=8cm,bb=0 0 792 603]{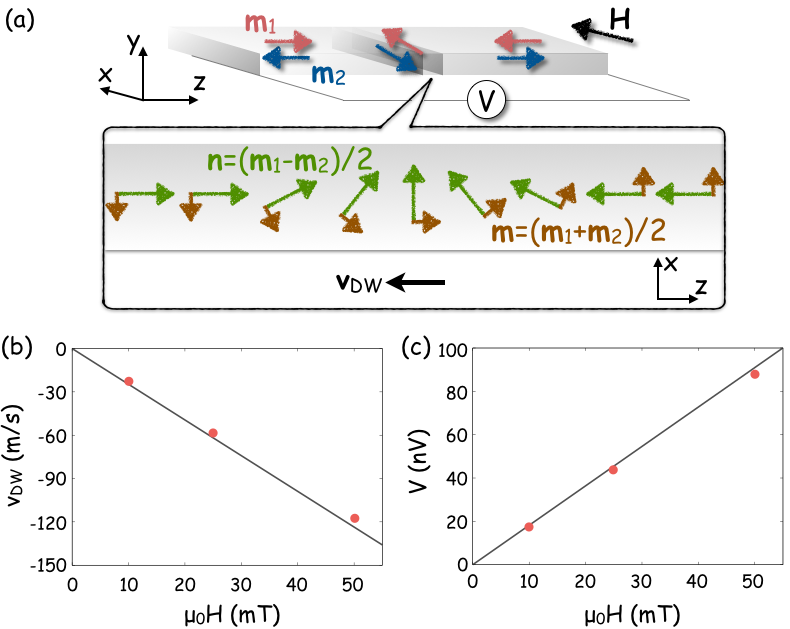}
\caption{ (a) Schematic of the present system where SMF is induced by one-dimensional DW motion.
See the main text for the definitions of the symbols.
(b) The DW velocity and (c) the electric voltage between the wire edges as functions of the applied magnetic field.
The solid lines plot Eqs.~(\ref{velocity}) and (\ref{v_dw}), while the red circles are the results of numerical simulations. }
\end{figure}

The electric voltage $V$ induced between the wire edges due to the DW motion is computed by Eq.~(\ref{v}) as
\begin{equation}
V \simeq \frac{P\beta}{2e} \int_{-\infty}^\infty dz \frac{\partial{\bm n}}{\partial t} \cdot \frac{\partial{\bm n}}{\partial z}
= - \frac{\hbar P \beta}{e\Delta} v_{\rm DW} .
\label{v_dw}\end{equation}
Fig.~1~(b) and (c) compare the analytical results of Eqs.~(\ref{velocity}) and (\ref{v_dw}), respectively, with direct numerical simulations based on Eq.~(\ref{e}) and (\ref{llg}), showing very good agreement.
Assuming $|t_{12}|/J=0.1$, the other parameters have been chosen so that they are in the typical range for AFM materials\cite{text}; $\gamma=1.76\times10^{11}$ Hz/T, $\mu_0M_{\rm S}=1$ T, $\mu_0H_E=40$ T, $A_1=1.6\times10^{-12}$ J/m, $K=2\times10^4$ J/m$^3$, $D=2\times10^6$ J/m$^3$, $\alpha=0.01$, $\beta=0.02$, and $P=0.5$.
In the simulations, Eq.~(\ref{llg}) is solved in a two-dimensional nanowire with dimensions of $2000\times20$ nm$^2$, dividing it into the unit cells of $2\times2$ nm$^2$.
The Poisson equation $\nabla\cdot\sum_i P \bm{{\cal E}}_+^i/2=\nabla^2\phi$ is solved self-consistently to obtain the spatial-distribution of $\phi$\cite{ohe,yamane-scirep}.
The small deviation of the full numerical results from the analytical lines for higher fields is mainly attributed to the approximation used in Eq.~(\ref{velocity}) for the magnetization dynamics.
We confirmed that the contribution from ${\bm f}({\bm n},{\bm m})$ in Eq.~(\ref{e_sum}) is negligibly small in the present case.

It is known that when the DW velocity reaches values as high as the sound velocities in the medium, the impact of the coupling of the DW with phonons becomes significant, leading to highly nonlinear behaviours of the DW\cite{bary}.
The maximum value of the DW velocity in AFMs is dictated by the magnon dispersion, and can be as high as of the order of $10$ km/sec\cite{bary}.
We leave for future work the systematic study of the SMF in such nonlinear regimes.
We conclude this part by pointing out the possibility of real-time observation and detailed investigation of the DW dynamics at a wide range of velocities by electrical means using SMF.

%=========================================================================
%          AFMR
%=========================================================================
\emph{Antiferromagnetic resonance.---}
In this part, we discuss SMF produced by a uniform AFMR in a thin film with RSOC, where the relevant term in Eq.~(\ref{e_sum}) is either the second or third term.

The system we consider is schematically shown in Fig.~2~(a).
The AFM thin film possesses the uniaxial magnetic anisotropy in the $y$ axis, and no DM interaction;
$u=-2\mu_0 {\bm m} \cdot {\bm H} + A_0 {\bm m}^2 - K  ( n_y^2 + m_y^2 ) $, where the neighboring exchange coupling has been omitted as we consider a uniform dynamics here.
The external dc field $H_{\rm dc}$ along the $y$ axis and the microwave with angular frequency $\omega$ excite AFMR, where ${\bm m}_1$ and ${\bm m}_2$ precess with the same angular frequency $\omega$, making the angles $\theta_1$ and $\theta_2$, respectively, with respect to their equilibrium directions.
The relative angle of ${\bm m}_1$ and ${\bm m}_2$ in the $x$-$z$ plane can be assumed to be always $\pi$ due to the mostly dominant AFM exchange coupling\cite{text}.

Because of the uniformity in the dynamical magnetic profile, in Eq.~(\ref{e_sum}) the terms that contain $\nabla{\bm m}$ and $\nabla{\bm n}$ do not come into play.
The electric voltage $V$ in the $y$ direction is obtained from Eq.~(\ref{v}) as
\begin{equation}
V = - \frac{ m_{\rm e} P }{2e\hbar}  L_y  \left( \lambda_1 \sin\theta_1 - \lambda_2 \sin\theta_2 \right) \omega \sin\omega t ,
\label{v_afmr}\end{equation}
where $L_y$ stands for the sample lengths in the $y$ direction.
We emphasize here that the amplitude of the ac electric voltage can be controlled by sample dimensions.

The values of $\omega$, $\theta_1$ and $\theta_2$ at a resonance condition can be estimated based on  Eq.~(\ref{llg}) as $\omega/\gamma= H_{\rm dc}+\sqrt{H_A ( H_A + 2 H_E )}$, $\theta_1=H_{\rm ac}\sqrt{\omega - \gamma (H_{\rm dc} - H_A )}/4\alpha H_E\sqrt{\omega}$, and $\theta_2= H_{\rm ac} \sqrt{\omega-\gamma (H_{\rm dc} + H_A )} /4\alpha H_E\sqrt{\omega}$\cite{text}, with $H_{\rm ac}$ the amplitude of the ac field corresponding to the microwave, and $H_A=K/\mu_0M_{\rm S}$.
Here $m_{1y}\simeq-m_{2y}\simeq1$, $\alpha\ll1$ and $H_E\gg H_{\rm dc},H_A$ have been assumed.

Fig.~2 (b) [(c)] plots the time evolution of Eq.~(\ref{v_afmr}) with $\lambda_1=\lambda_2=10^{-10}$ eV$\cdot$m [$\lambda_1=-\lambda_2=10^{-10}$ eV$\cdot$m\cite{note3}].
We have employed $\mu_0 H_A=0.6$ T, $\mu_0H_{\rm dc}=4.4$ T, $\mu_0H_{\rm ac}=0.2$ mT, $L_y=10$ $\mu$m, and the same values for the other parameters as before.
These values give $\omega\simeq2$~THz$\cdot$rad, $\theta_1\simeq0.0047^\circ$ and $\theta_2\simeq0.0039^\circ$.
The larger amplitude of $V$ in the case of $\lambda_1=-\lambda_2$ reflects the condition $|{\bm n}_{\rm neq}|>|{\bm m}|$, where ${\bm n}_{\rm neq}$ is the non-equilibrium, i.e., oscillating, component of ${\bm n}$.

\begin{figure}
\centering
\includegraphics[width=8cm,bb=0 0 775 625]{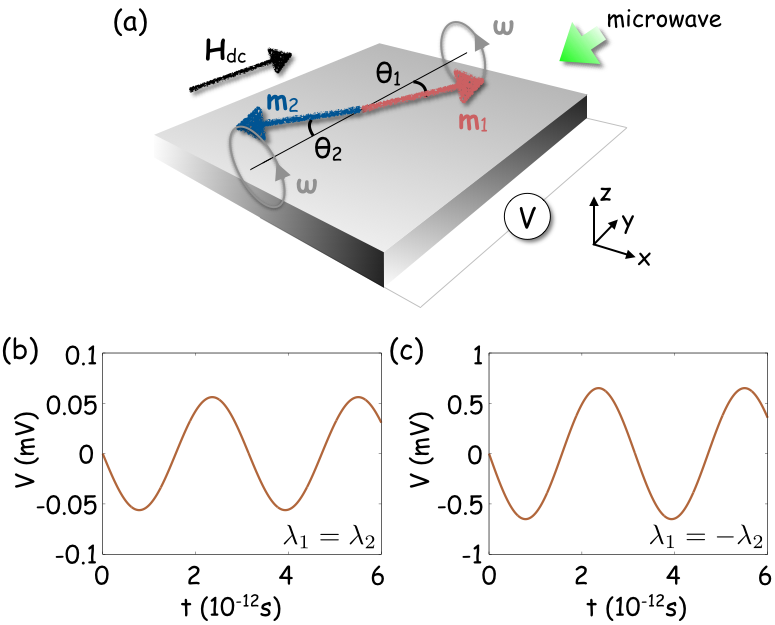}
\caption{(a) Schematic of the present system of generating SMF by AFMR.
See the main text for the definitions of the symbols.
(b),(c) Time dependence of the electric voltage $V$ in Eq.~(\ref{v_afmr}) for the case of (b) $\lambda_1=\lambda_2$ and (c) $\lambda_1=-\lambda_2$. 
 }
\end{figure}

%=========================================================================
%           Conclusions%=========================================================================
In conclusion, we have predicted SMF in two-sublattice AFMs, demonstrating dc and ac electric voltage generation in the systems that involve DW motion and AFMR, respectively.
Our results indicate that SMF can play an important role in the antiferromagnetic spintronics as it offers a way to electrically detect the dynamical AFM textures.

%=========================================================================
%           Acknowledgments
%=========================================================================
\emph{Acknowledgments.---}
The authors are grateful to H. Gomonay and R. Cheng for valuable comments on the manuscript, and K. Yamamoto for fruitful discussions.
This research was supported by Research Fellowship for Young Scientists from Japan Society for the Promotion of Science, Grant-in-Aid for Scientific Research (No.~24740247, 26247063) from MEXT, Japan, and Alexander von Humboldt Foundation.
%=========================================================================
%           References
%=========================================================================


\begin{thebibliography}{99}
\bibitem{note4} The explicit forms of $t_{11}$ and $t_{12}$ reflect the actual lattice structure and the nature of atomistic electron hopping. 
For example, $t_{11}= t_{11}(p_x,p_y)$ and $t_{12}=t_{12}(p_z)$ in a layered AMF where FM and AFM orderings are realized in the $x$-$y$ plane and along the $z$ axis, respectively.
In the nearly-free electron model, they may be written as $t_{11}=(p_x^2+p_y^2)/2m_{\rm e}$ and $t_{12}=p_z^2/2M_{\rm e}$ with different effective masses $m_{\rm e}$ and $M_{\rm e}$. 
When the inter-layer hopping, i.e., inter-sublattice dynamics is relatively suppressed with $M_{\rm e}\gg m_{\rm e}$, then $t_{12}$ may be approximated by its value at the Fermi surface.
For our purpose, the constant approximation for $t_{12}$ is enough as it does not play an active role.
\bibitem{note} The last term in Eq.~(\ref{off}) is apparently of the order of $J^{0}$.
However, ${\bm m}\propto J_{\rm sub}^{-1}$\cite{andreev}, where $J_{\rm sub}$ is the AFM exchange coupling energy between ${\bm m}_1$ and ${\bm m}_2$, which is as large as $J$.
Therefore, $L$ is of the order of the inverse of the exchange couplings.
\bibitem{note3}  The case with $\lambda_1 = - \lambda_2$ corresponds to a system where the two sublattices individually have broken inversion symmetry and form inversion partners\cite{jakob}.
In those systems each sublattice provides opposite inverse spin galvanic effects, such as staggered spin-orbit torques.
\end{thebibliography}
\end{document}